\documentstyle[mncite]{mn}
\title[Distance measurement of nearby open clusters]
{A new method for estimating the distance of young open clusters}

\author[M.A. Hendry, M.A O'Dell and A. Collier Cameron]
        {M.A. Hendry, M.A. O'Dell and A. Collier Cameron\\
         Astronomy Centre, University of Sussex,
         Falmer, Brighton BN1 9QH, UK.}

\date{Accepted 1993 June 30. Received 1993 June 7;
in original form 1993 May 31}

\begin{document}

\maketitle

\newcommand{\msc}[1]{\mbox{\sc #1}}
\newcommand{\rmsub}[2]{#1_{\rm #2}}
\newcommand{\bu}[1]{\bf{\underline {#1}}}

\begin{abstract}

We present a new technique for estimating the distance to
young open clusters. The method requires accurate measurement of the
axial rotation period of late-type members of the cluster: rotation periods are
first combined with projected rotation velocities and an estimate of the
angular diameter for each star - obtained using the Barnes-Evans relation
between colour and surface brightness. A `best' cluster distance estimate is
then determined using standard techniques from the theory of order statistics,
which are in common use in the general statistics literature.
It is hoped that this new method will prove a useful adjoint to more
traditional distance methods, in order to better ascertain the distance scale
within the solar neighbourhood.
\end{abstract}


\noindent

\section{INTRODUCTION}
\label{sec:intro}

The accurate measurement of distances on intergalactic and extragalactic
scales poses a problem of critical importance to both astronomical and
cosmological research.
There exists in common use a variety of different distance indicators;
bridging and relating these indicators has enabled the construction of a
cosmological distance ladder, the precise calibration of which is heavily
reliant upon several key distances.

Of fundamental importance is the distance determination of nearby galactic
clusters: for example the Hyades distance is a crucial `yardstick' upon
which the calibration of the extragalactic distance scale and ultimately the
Hubble constant depend.

In this paper we describe a new technique for estimating the distance to nearby
galactic clusters. The method is applicable to young clusters containing fast
rotating late type stars, the periods of which may be determined by
measuring rotational modulation due to surface inhomogeneities.
Prime examples of such clusters are the Pleiades and $\alpha$
Persei, which are the next step beyond the Hyades in the standard
zero-age main sequence (ZAMS) fitting procedure.

The idea of using stellar rotation periods to estimate cluster distances was
previously discussed in \scite{cameron92alpper} (hereafter CCW)
although no detailed statistical analysis was attempted. In CCW it was observed
that the measured rotation period, $P$, of a star may be combined with the
projected rotational velocity, $v \sin i$, to determine an estimate of the
star's `projected' radius, $R \sin i$, viz:-
\begin{equation}
R \sin i = \frac{ P v \sin i }{ 2 \pi }
\end{equation}
Here the inclination, $i$, takes its usual definition: the angle between the
rotation axis of the star and the line of sight.

In CCW an estimate of the stellar angular diameter, $\phi$, was then inferred
for each star using the relation between colour and surface brightness
derived by \scite{bem78} (hereafter BEM).( See also Section \ref{sec:bevans}
below). A `projected' cluster distance, $D \sin i$, then follows in the
obvious way, viz:-
\begin{equation}
D \sin i = \frac{ 2 R  \sin i }{ \phi }
\end{equation}
where $\phi$ is measured in radians. Substituting from equation (1) we find:-
\begin{equation}
D \sin i = \frac{ P v \sin i }{ \pi \phi }
\end{equation}
Or, expressing in more convenient units:-
\begin{equation}
D \sin i = 7.660 \times 10^{-3} \quad \frac{ P v \sin i }{ \phi }
\end{equation}
where now $P$ is measured in hours, $v \sin i$ in kms$^{-1}$, $\phi$ in
milliarcseconds and $D \sin i$ in parsecs. This is essentially equation (4) of
CCW.

$D \sin i$ values were thus derived in CCW for five stars in the $\alpha$ Per
cluster. The largest value of $D \sin i$ was adopted as a crude
estimate of the cluster distance: this was based on the assumption that the
largest observed $D \sin i$ should correspond to the star with the highest
axial
inclination, so that $ \sin i$ might reasonably be taken to be equal to unity
- the maximum value in the distribution of $ \sin i$.

This simple approach did indeed give a cluster distance estimate which was in
reasonable agreement, but greater than, the distance modulus $(m - M)_o =
6.1$ quoted in \scite{stauffer85}. The limitations of the approach, however,
stem from the assumption that the distribution of $D \sin i$ values arises
solely
from the intrinsic distribution of stellar inclinations, when in practice it
results from a number of contributory factors: the intrinsic distribution of
rotation periods, rotation velocities, inclinations and diameters - together
with the observational scatter in the measurement of each of these quantities.

The aim of this paper is to extend and improve the simple treatment of CCW by
addressing these limitations, and thus place the estimation of cluster
distances
from stellar rotation periods on a more rigorous statistical footing. In
Section \ref{sec:bevans} we begin by describing in more detail the principles
behind the calibration and use of the Barnes-Evans relation, upon which our
distance method
relies. In Section \ref{sec:dsini} we then carefully model the intrinsic
scatter and observational selection effects which contribute to the
distribution of $D \sin i$ values. Such a detailed treatment is essential
before one can meaningfully assign an error to any cluster distance estimate.

Having thus derived the $D \sin i$ distribution expected for a given cluster,
in
Section \ref{sec:ests} we consider several different distance estimators
which one may define in terms of the observed $D \sin i$ values. We
investigate the properties of
these estimators, demonstrating how one may assign the appropriate distance
error to each, and how they may be used to construct confidence intervals for
the true cluster distance. In particular we show how one may improve
upon the crude estimator adopted in CCW by combining the $D \sin i$
values inferred for {\em all} of the observed stars - instead of using simply
the largest value.
Finally, using artificially generated data, we examine how
the accuracy of our new distance method varies with the number of observed
cluster stars, and the intrinsic scatter in the Barnes-Evans relation. Based
on these results we assess the usefulness of our new method set
alongside the more traditional cluster distance indicators such as the ZAMS
fitting procedure. In a forthcoming paper, currently in preparation, we will
apply our distance method to a larger sample of $\alpha$ Per stars and the
Pleiades cluster for which rotation periods have been determined, and make a
direct quantitative comparison between the cluster distance estimated by our
method and that derived from ZAMS fitting.

\section{THE RELATION OF BARNES AND EVANS (Stellar angular diameters vs visual
surface brightness/colour)}
\label{sec:bevans}

In order to calculate a projected cluster distance, $D \sin i$,
we have seen in Section \ref{sec:intro} that it is necessary to estimate the
angular diameter $\phi$ of each star. This can be found to a limited degree of
accuracy by introducing the work of \scite{barnesevans76} (hereafter BE) who
correlated stellar angular diameters to visual surface brightness and thus to
colour index.

This correlation had been known for some time. Similar work was carried out
by \scite{wesselink69}, \scite{warner72} and \scite{harwood75} but was limited
basically to early type stars and the B-V colour index. The work of Barnes
and Evans extended the correlation to the entire UBVRI system including red
stars to spectral type M8. Their work was partly instigated by the sudden
availability of many more stellar angular diameters found through a program
observing the lunar occultations of nearby stars.

Barnes and Evans defined a quantity $\rmsub{F}{v}$ the visual surface
brightness parameter which can be shown to be:-
\begin{equation}
\rmsub{F}{v} = 4.2207 - 0.1V_o - 0.5log{\phi}
\end{equation}
where $V_o$ is the unreddened apparent magnitude and $\phi$ is the stellar
angular diameter expressed in milliseconds of arc.

Computing $\rmsub{F}{v}$ for the nearby calibration stars and then plotting the
resultant values against the various colour indices; B-V, V-R and R-I
(shown in BE) confirmed a linear relationship between these indices and
surface brightness. (A limitation was imposed for B-V where the relation
breaks down and the index no longer relates to the energy output of the star).
In BEM Barnes, Evans and Moffett improved these relations by adding more
stars and correcting for the effects of limb darkening.

Of all the relationships the V-R vs $\rmsub{F}{v}$ appears to have the tightest
correlation but for the purposes of this paper we will follow the work of
CCW and employ the B-V vs $\rmsub{F}{v}$ relation which is linear over the
spectral range of the $\alpha$ Per G-K dwarfs.

{}From BEM the B-V vs $\rmsub{F}{v}$ relation is linear over the interval:-
\begin{equation}
\rmsub{F}{v} = 3.964 - 0.333(B-V)_o\;,\quad -0.10 {\leq} (B-V)_o {\leq} 1.35
\end{equation}
where $(B-V)_o$ is the unreddened colour index.

Equating equation (5) (theoretically derived) to equation (6) (empirically
derived) and converting to reddened colour index and
apparent magnitude gives the semi-empirical relation:-
\begin{equation}
log{\phi} = 0.5134 - 0.066E_{B-V} + 0.666(B-V) - 0.2V
\end{equation}
assuming ${A_v} \simeq 3.0E_{B-V}$ for the Perseus region
\scite{hiltner56} (see equation (3), CCW)

We can, therefore, obtain $\phi$ from a knowledge of B-V, $V$ and the colour
excess $E_{B-V}$ - for which an average value of 0.10 - 0.11 was adopted for
the $\alpha$ Per region (\scite{CrawBarn74} and \scite{Pross91}).

In effect we are working the original computation of Barnes and Evans
backward to obtain $\phi$ from $\rmsub{F}{v}$ given colour and apparent
magnitude $V$.

Most of the uncertainty in the stellar angular diameter $\phi$ stems from
the original lunar occultation measurements of the nearby calibration stars
(typically 5 - 20 \% (BE). More difficult to quantify (due to the paucity of
calibration stars) is the uncertainty in the interpolation of the
$\rmsub{F}{v}$ vs B-V graph within the range of our G-K dwarfs.
Finally to a lesser degree we must also consider
the systematic errors inherent from the measurement of B-V, $V$ and $E_{B-V}$
and the approximation $A_v \simeq 3.0E_{B-V}$.

The computed values of $\phi$ generated from equation (6) are then
substituted into equation (3) to obtain a $D \sin i$ value for each star whose
axial rotation period and projected equatorial velocity are known.
Having generated a range of $D \sin i$ values we then obtain a number of
different cluster distance estimates by interpreting the `observed'
distribution of $D \sin i$ values according to the statistical methods
described in Section \ref{sec:ests}.

\section{DISTRIBUTION OF Dsini}
\label{sec:dsini}

In this section we construct a model for the distribution of $D \sin i$ values
which one would expect to observe in a cluster at a given true distance. We
will then later use this distribution as our basis for defining different
cluster distance estimators and studying their properties.

A general method for deriving this distribution may be found in any elementary
statistics textbook. Firstly one adopts a model for the joint probability
distribution of the observed rotation period, rotation velocity, inclination
and
angular diameter - the variables in terms of which $D \sin i$ is defined.
This distribution must take account of the intrinsic spread in these
variables  from star to star, the scatter due to measurement error in the
values observed for a given star, and any observational selection effects to
which the sample may be subject. One next uses equation (3) to define an
appropriate transformation of $P$, $v$, $i$ and $\phi$, such that the
distribution of $D \sin i$ may then be extracted directly from the joint
distribution of the transformed variables.

Although this method is straightforward to apply in principle, it rarely
yields an analytic solution in practice, and that is indeed the case here.
The problem is easily tackled via Monte Carlo simulations, however: i.e. we
draw a large number of sets of the random variables, $P$, $v$, $i$ and $\phi$,
and for each set compute the corresponding value of $D \sin i$. We then
deduce the
distribution of $D \sin i$ by constructing a histogram of the computed values.

Our analysis can be further simplified by making one additional assumption
and a judicious change of variables, as we now show.

Suppose that we observe a cluster which lies at true distance $\rmsub{D}{true}$
parsecs. We can regard $\rmsub{D}{true}$ as a fixed - though of course unknown
-
parameter, which we wish to estimate. Consider a star in the cluster and let
$P_{true}$, $v_{true}$ and $\phi_{true}$ denote the true
rotation period, rotation velocity and angular diameter (in radians)
respectively of this star. Observe that these variables are not mutually
independent, but are related by the following equation:-
\begin{equation}
\rmsub{D}{true} = \frac{\rmsub{P}{true} \, \rmsub{v}{true}}
{\pi \, \rmsub{\phi}{true}}
\end{equation}
Equation (8) assumes that the star lies precisely at the centre of the cluster
-
i.e. we neglect any line of sight depth.

Consider again equation (3) above, which expresses $D \sin i$ in terms
of the {\em observed} values of $P$, $v$, $i$ and $\phi$. Introducing two new
variables, $\rmsub{z}{p} = \frac{\rmsub{P}{obs}}{\rmsub{P}{true}}$, and
$z_{\phi} = \frac{\rmsub{\phi}{obs}}{\rmsub{\phi}{true}}$, and combining with
equations (3) and (8) we may write:-
\begin{equation}
D \sin i  =  \frac{\rmsub{D}{true} \: \rmsub{(v \sin i)}{obs} \: \rmsub{z}{p}}
{\rmsub{v}{true} \: z_{\phi}}
\quad \quad \equiv \quad \rmsub{D}{true} \: \alpha
\end{equation}
This change of variables has two immediate advantages. Firstly, we no longer
require to model the {\em intrinsic} distribution of $P$ and $\phi$, only
the observational scatter about their true values. One might reasonably
expect these error distributions to be independent of the value of
$\rmsub{P}{true}$ and $\rmsub{\phi}{true}$ respectively. Secondly, and more
importantly, introducing
$\rmsub{z}{p}$ and $z_{\phi}$ simplifies the dependence of the $D \sin i$
distribution upon $\rmsub{D}{true}$. Clearly once we have modelled the
distribution of the composite variable, $\alpha$, in equation (9), we can
obtain the distribution of $D \sin i$ for a cluster at {\em any} true
distance simply by rescaling. One may
think of $\alpha$ as a dimensionless projection factor, essentially equivalent
to $\sin i$.

As an example, Figures \ref{fig:alpha5pc} and \ref{fig:alpha10pc} show
probability density curves for $\alpha$, determined for a particular
selection function, and for different angular diameter error dispersions.
The probability density curves were obtained by spline fitting to histograms
constructed from 50000 trials. The Monte Carlo sampling was carried out as
follows.

\begin{enumerate}
\item The true rotation velocity was first drawn from a uniform distribution
in the range 0 to 240kms$^{-1}$. For a star of one solar radius an equatorial
velocity of 240kms$^{-1}$ corresponds to a rotational period of $\sim$ 5
hours, which was the shortest measured rotation period in the $\alpha$ Per
sample studied in \scite{odell93}.

\item A true inclination was then assigned, based on the standard assumption
that the orientation of the stellar rotation axis is completely random with
no preferred direction in space (\scite{slettebak70}). It follows
easily from this assumption that the {\em intrinsic} distribution of $cosi$
is uniform over the interval $[0,1]$.

\item The observed $v \sin i$ was assigned by multiplying
$\rmsub{(v \sin i)}{true}$
by a Gaussian of unit mean and dispersion of 0.1; a relative error of $10\% \,$
being typical for measurements of spectral line broadening of cluster stars
(\scite{stauffer85}).

\item A lower selection limit of
$\rmsub{(v \sin i)}{obs} \geq$ 50kms$^{-1}$ was then
imposed. This was the limit adopted in selecting the $\alpha$ Per sample
studied in \scite{odell93}. This selection function was primarily
designed to ensure that only fast rotators were included in the sample  - thus
improving the chances of detecting rotational modulation within the available
observing time.

\item Finally, the scaled variable, $z_{\phi}$ was drawn from a gaussian of
unit
mean and constant dispersion, $\sigma_{\phi}$, where $\sigma_{\phi} = 0.05$ in
Figure \ref{fig:alpha5pc}, and $\sigma_{\phi} = 0.1$ in Figure
\ref{fig:alpha10pc}. Note that assigning a
constant {\em percentage} error dispersion to the observed angular diameter
was equivalent to a gaussian scatter of constant dispersion in the linear
Barnes-Evans relation for $log \phi$, as given by equation (7).

\item In all cases the other scaled variable, $\rmsub{z}{p}$, was set
identically equal to unity: i.e. it was assumed that the true axial rotation
period of each star was recovered exactly from fourier analysis of the light
curve. This
approximation seemed reasonable when the data quality was sufficiently high to
ensure little ambiguity in the star's power spectrum. (See
\scite{odell93} for further discussion). In such a case the error in the
derived period would be considerably smaller than the uncertainty in both
$\rmsub{\phi}{obs}$ and $\rmsub{(v \sin i)}{obs}$.
\end{enumerate}
\begin{figure}
  \centering
  \vspace{6cm}
\caption{Probability density function of the scaled variable, $\alpha$,
derived from a spline fit to the histogram obtained from $50000$ Monte Carlo
trials. The dispersion of the angular diameter errors was assumed to be
$\sigma_{\phi} = 0.05$. The dashed curve indicates the intrinsic distribution
of $\rmsub{\sin i}{true}$ in the absence of observational errors and selection
effects.}
\label{fig:alpha5pc}
\end{figure}
\begin{figure}
  \centering
  \vspace{6cm}
\caption{Probability density function of the scaled variable, $\alpha$,
derived from a spline fit to the histogram obtained from $50000$ Monte Carlo
trials. The dispersion of the angular diameter errors was assumed to be
$\sigma_{\phi} = 0.1$. The dashed curve indicates the intrinsic distribution
of $\rmsub{\sin i}{true}$ in the absence of observational errors and selection
effects.}
\label{fig:alpha10pc}
\end{figure}
It is important to note at this point that the lower selection limit,
$\rmsub{(v \sin i)}{obs} \geq$ 50kms$^{-1}$, also serves indirectly to exclude
stars of low inclination ($i < 12^{\circ} \,$), since we are imposing an
{\em upper} limit on the true rotation velocity. We would certainly expect some
selection of this kind to exist in our sample, since stars viewed at or near to
pole-on would not display significant rotational modulation, and hence their
periods could not be measured photometrically. The exact form of this
selection at higher inclinations will depend upon the distribution of surface
inhomogeneities, and in a more rigorous treatment this dependence could be
modelled explicitly in our Monte Carlo sampling: i.e. for a given surface
distribution we could determine the probability of detecting rotational
modulation as a function of inclination. We do not incorporate such
a model in this paper, however, since our aim is primarily to
illustrate the essential features of our new distance method. Thus we adopt an
inclination selection given simply by the limit of $i < 12^{\circ} \,$,
as indicated above. In our next paper we will consider the precise form of
the inclination selection in more detail - together with the inverse problem
of how one might use our cluster distance estimate to infer the
distribution of surface inhomogeneities for the sampled stars.

We can see from equation (9) that, if one has perfect, selection-free,
measurements of $v \sin i$ and $\phi$ then $\alpha$ is identically
distributed as $\rmsub{\sin i}{true}$. Hence, in this ideal case, the $\alpha$
distribution rises monotonically from $\alpha = 0$, peaks at $\alpha = 1$ and
drops immediately to zero for all $\alpha > 1$, as shown by the dashed curves
in Figures \ref{fig:alpha5pc} and \ref{fig:alpha10pc}. Recall from section
\ref{sec:intro} that this behaviour motivated the choice in CCW of the
largest $D \sin i$ as the cluster distance estimate. When we include the
effects of measurement errors and observational selection on $v \sin i$ and
$\phi$, however, we find that the {\em observed} distribution of $\alpha$ - as
indicated by the solid curves - is substantially different from the intrinsic
distribution of $\rmsub{\sin i}{true}$, and this difference reveals a serious
weakness in the CCW approach. Firstly we can see that the $\alpha$
distribution no longer extends to $\alpha = 0$ - i.e. the sample is biased
against stars of low inclination, as discussed above. More importantly,
however,
the distribution displays a significant tail for $\alpha > 1$.
Specifically, when $\sigma_{\phi} = 0.05$ approximately $20\% \, $ of the
$\alpha$ distribution lies in the range $\alpha \geq 1$. From equation (9),
therefore, the probability equals $\sim 0.2$ that one would infer - for any
given star - a $D \sin i$ greater than the true cluster distance. This
probability increases to $\sim 0.25$ for $\sigma_{\phi} = 0.1$. Moreover, for
the sample of five stars considered in CCW, the probability that the
{\em largest} value of $D \sin i$ is greater than $\rmsub{D}{true}$ increases
to $\sim 0.68$ for $\sigma_{\phi} = 0.05$, and to $\sim 0.75$ for
$\sigma_{\phi} = 0.1$. (These numbers follow easily from standard results on
the distribution of {\em order statistics}, which we will introduce in
Section \ref{sec:ests}, below). Thus we see that the cluster distance derived
in CCW is quite likely an {\em over} - estimate, as a result of the failure
to model the errors and selection on $v \sin i$ and $\phi$. Note that the CCW
estimate of the $\alpha$ Per distance modulus was indeed somewhat larger than
that quoted in e.g. \scite{stauffer85}.

Can one define a better estimate of the cluster distance? Having shown how
one may model the $D \sin i$ distribution for any given cluster, we can now
provide a quantitative answer to this question.

\section{CLUSTER DISTANCE ESTIMATORS}
\label{sec:ests}

Given a set of $D \sin i$ values inferred from our sample of cluster stars we
may
define a number of different cluster distance estimators by combining these
$D \sin i$ values in various different ways. In this section we define and
compare
the properties of four such estimators, beginning with the estimator adopted in
CCW.

\subsection{`Naive' Estimator, $\rmsub{\hat{D}}{naive}$}

This estimator is simply the largest value of $D \sin i$ in our sample, as
discussed previously. Since we ignore the effects of errors in $v \sin i$
and $\phi$ in defining this estimator, we refer to it as `naive'.
Suppose we observe a sample of $n$ stars, and that we order the $D \sin i$
values
inferred for these stars in increasing size. Let $\{ D \sin i_{(k)}
\hspace{2mm} ;
k = 1 , n \} $ denote the ordered sample. i.e.:-
\begin{equation}
D \sin i_{(1)} \leq D \sin i_{(2)} \leq \ldots \leq D \sin i_{(n)}
\end{equation}
Then $\rmsub{\hat{D}}{naive}$ is defined simply as:-
\begin{equation}
\rmsub{\hat{D}}{naive} = D \sin i_{(n)}
\end{equation}
Note that we are adopting the usual statistical convention of using a caret
to denote an estimator of a parameter. We will return to the statistical
properties of ordered samples later in this section.

\subsection{`Mean' Estimator, $\rmsub{\hat{D}}{mean}$}

If we are to define a better distance estimator than $\rmsub{\hat{D}}{naive}$
then clearly we must take some account of the true shape of the $D \sin i$
distribution. One might also wish to make use of all the stars observed in
the cluster, instead of just the star with the largest $D \sin i$. One
obvious way to do this is by using the mean $D \sin i$ of our sample.

Consider again equation (9) above. If we take the mean of both sides (more
formally, the expectation value over the $D \sin i$ and $\alpha$ distributions
at fixed true distance, $\rmsub{D}{true}$), then it follows trivially that:-
\begin{equation}
< \hspace{-0.7mm} D \sin i \hspace{-0.7mm} > = \rmsub{D}{true}
\hspace{0.2mm} < \hspace{-0.7mm} \alpha \hspace{-0.7mm} >
\end{equation}
We thus define our estimator, $\rmsub{\hat{D}}{mean}$, in terms of
the sample mean value of $D \sin i$ and $< \hspace{-0.7mm} \alpha
\hspace{-0.7mm} >$,
the mean value of $\alpha$ as determined from our modelled distribution
function, viz:-
\begin{equation}
\rmsub{\hat{D}}{ mean} = \frac{ 1 }{ n \hspace{0.1mm} < \hspace{-0.7mm} \alpha
\hspace{-0.7mm} > } \sum_{k=1}^{n} D \sin i_{(k)}
\end{equation}
An equivalent way of looking at this estimator is as follows: for each star
individually we can derive a distance estimate, which is given by the cluster
distance required so that the inferred $D \sin i$ for that particular star is
equal to the mean value of $D \sin i$ at that distance.
$\rmsub{\hat{D}}{mean}$ is then simply equal to the mean value of
these individual distance estimates.

\subsection{`Median' Estimator, $\rmsub{\hat{D}}{med}$}

We can derive another estimator based on the form of the $\alpha$
distribution - this time using the median value, $\rmsub{[\alpha]}{med}$.
Since this distribution is asymmetric, as we can see from
Figures \ref{fig:alpha5pc} and \ref{fig:alpha10pc},
$\rmsub{[\alpha]}{med}$ will not generally equal the mean value,
$< \hspace{-0.7mm} \alpha \hspace{-0.7mm} >$. Our distance estimate is
defined as
the true distance for which the median of our ordered sample of observed
$D \sin i$ values is equal to the the median of the expected $D \sin i$
distribution at that distance. Hence we have:-
\begin{equation}
\rmsub{\hat{D}}{med} = \frac{\rmsub{[D \sin i]}{med}}
{\rmsub{[\alpha]}{med}}
\end{equation}
where $\rmsub{[\alpha]}{med}$ denotes the median of the modelled $\alpha$
distribution and $\rmsub{(D \sin i)}{med}$ denotes the median of the sampled
$D \sin i$ values.
We can write down an expression for $\rmsub{[D \sin i]}{med}$ in terms of the
elements of our ordered sample, viz:-
\begin{eqnarray}
\rmsub{[D \sin i]}{med} = \left\{ \begin{array}{ll}
                      D \sin i_{(\frac{n+1}{2})}  & \hspace{1mm}
                                      \mbox{$n$ odd} \nonumber \\
                      \frac{1}{2}(D \sin i_{(\frac{n}{2})} +
                                  D \sin i_{(\frac{n}{2}+1)})
                      & \hspace{1mm} \mbox{$n$ even} \nonumber \\
                  \end{array}
                  \right.
\end{eqnarray}
We can use a Monte Carlo approach to deduce the distribution of each of these
estimators, and hence compare their relative accuracy. We generate a large
number of artificial cluster samples of a given size - each sample drawn from
our modelled $D \sin i$ distribution for a cluster at some chosen true
distance.
For each sample we then compute the cluster distance estimate predicted by each
of the three estimators defined above. We determine the distribution of each
estimator at the specified true distance by constructing a histogram of the
computed values.

As an illustration, Figures \ref{fig:est10stars} to \ref{fig:est30stars}
show a range of results obtained from
generating 2000 cluster samples - containing $10, 20$ and $30$ stars, and for
angular diameter error dispersions of $5\% \, $ and $10\% \, $ respectively.
In all cases the true cluster distance was taken to be $200 \,$pc, as indicated
by the dashed line on each histogram. The mean values and dispersions
of each estimator, as calculated directly from the histograms, are summarised
in Table \ref{tab:estdistrib}.

A consistent picture of the properties of the estimators emerges from these
results. In all cases the naive estimator is positively {\em biased} - i.e.
the mean value of the estimator is systematically greater than the true
distance of $200 \,$pc. One would expect, therefore, $\rmsub{\hat{D}}{naive}$
to
systematically over-estimate the cluster distance, as was suggested in section
(3). Contrary to what one might first expect, the magnitude of the bias of
$\rmsub{\hat{D}}{naive}$ in fact {\em increases} as the sample size increases:
this is because, with a larger number of sampled stars, it becomes more likely
that one will sample a star which lies further into the positive tail of the
$\alpha$ distribution. Hence $\rmsub{\hat{D}}{naive}$ would become a
progressively
{\em poorer} distance estimator as we add more stars to our cluster sample.
In the most extreme example considered here, for a cluster sample of 30 stars
and for $\sigma_{\phi} = 0.1$, the mean value of $\rmsub{\hat{D}}{naive}$ is
more than $250 \,$pc - representing a positive bias of $\sim 25\% \, $. It
follows from the form of equation (9) that the percentage bias of
$\rmsub{\hat{D}}{naive}$ is independent of $\rmsub{D}{true}$, for a given
sample size, so
this figure would hold good for a cluster at any true distance.
The bias of $\rmsub{\hat{D}}{naive}$ also increases roughly in proportion to
the
angular diameter dispersion, $\sigma_{\phi}$, which is as one might expect.

The bias of $\rmsub{\hat{D}}{mean}$ and $\rmsub{\hat{D}}{med}$ on the other
hand is
negligible, irrespective of the sample size, for the cases considered here.
This clearly demonstrates the importance of properly accounting for the true
form of the $D \sin i$ distribution in defining a good cluster distance
estimate. From Table \ref{tab:estdistrib} we also see that the dispersion of
$\rmsub{\hat{D}}{mean}$ is
found to be slightly smaller than that of $\rmsub{\hat{D}}{med}$ in all cases,
so that the former is marginally the more accurate estimator. We defer further
discussion of Table \ref{tab:estdistrib} until later.

Having shown from our simulations that both $\rmsub{\hat{D}}{mean}$ and
$\rmsub{\hat{D}}{med}$
are better distance estimators than $\rmsub{\hat{D}}{naive}$, we now consider
whether one may improve still further upon the former two estimators.

In the case of a normal random variable, the sample mean is an example of
a {\em sufficient} statistic (c.f. \scite{MoodGray74}). Broadly speaking,
this means that if one wishes to estimate the mean, $\theta$, of the normal
distribution from a sample, $\{ x_{i}; \hspace{0.2mm} i = 1, \ldots, n \}$,
then the sample mean, $< \hspace{-0.7mm} x \hspace{-0.7mm} > = \frac{1}{n}
\sum x_{i}$, provides
precisely the same statistical information about $\theta$ as does the complete
set of sampled values $\{ x_{i} \}$. i.e. knowledge of each $x_{i}$ would not
improve our estimate of $\theta$ compared with that obtained from knowledge
of $< \hspace{-0.7mm} x \hspace{-0.7mm} >$ alone.
\begin{figure*}
  \centering
  \vbox to 165mm{}
\caption{Histograms of naive, mean and median cluster distance
estimator distributions, derived from 2000 samples of 10 stars at a true
distance of $200 \,$pc. Results are shown for
for an angular diameter error dispersion of $\sigma_{\phi} = 0.05$ and
$0.1$}
\label{fig:est10stars}
\end{figure*}
This property does not hold in general for random variables which are not
normally distributed, however. We can clearly see from Figures
\ref{fig:alpha5pc} and \ref{fig:alpha10pc}
that the distribution of $\alpha$ is highly non-Gaussian: one might expect,
therefore, that knowledge of the individual $D \sin i$ values in our sample -
and in particular their {\em ordering} - would allow one to define a `better'
(in the sense of having a smaller dispersion) distance estimator than
$\rmsub{\hat{D}}{mean}$ or $\rmsub{\hat{D}}{med}$. Before we introduce such
an estimator we first make some preliminary remarks about the properties of
order statistics.

The use of order statistics is a common technique in applied statistics, and
the subject is treated extensively in a number of textbooks and monographs
(c.f. \scite{David81}, \scite{Gumbel58}). Suppose we draw a sample of size n
from the distribution of our random variable, $\alpha$. By the
$r^{th}$ order statistic, which we denote by $\alpha_{(r)}$, we mean simply
the $r^{th}$ smallest member of the sample. Hence $\alpha_{(n)}$ denotes
the largest sampled value: this is precisely the notation introduced in
equation (10) above.

The probability density distribution, $p_{r}(\alpha_{(r)})$, of the $r^{th}$
order statistic is closely related to the distribution of the parent random
variable, $\alpha$, viz:-
\begin{equation}
p_{r}(\alpha_{(r)}) = \frac{n!}{(r-1)!(n-r)!} [P(\alpha_{(r)})]^{r-1}
                        [1 - P(\alpha_{(r)})]^{n-r} p(\alpha_{(r)})
\end{equation}
where $p(\alpha_{(r)})$ and $P(\alpha_{(r)})$ denote respectively the
probability density and cumulative distribution functions of the parent
random variable, $\alpha$, in both cases evaluated at $\alpha = \alpha_{(r)}$.
\begin{figure*}
  \centering
  \vbox to 165mm{}
\caption{Histograms of naive, mean and median cluster distance
estimator distributions, derived from 2000 samples of 20 stars at a true
distance of $200 \,$pc. Results are shown for
for an angular diameter error dispersion of $\sigma_{\phi} = 0.05$ and
$0.1$}
\label{fig:est20stars}
\end{figure*}
Figure \ref{fig:ordstats} illustrates the distribution of several of the
$\alpha$ order
statistics, for a sample of 10 stars drawn from the $\alpha$ distribution
shown in Figure \ref{fig:alpha5pc} - i.e. for $\sigma_{\phi} = 0.05$. We can
understand
qualitatively how the shape of these distributions emerges from the form
of equation (16). The distribution of $\alpha_{(1)}$ for example has a larger
dispersion than that of the higher order statistics
$\alpha_{(7)}$ and $\alpha_{(10)}$ shown here. This is because the shape of
$p_{1}(\alpha_{(1)})$ is determined primarily by the behaviour of
$p(\alpha)$ at small $\alpha$, in which range the term $[1 - P(\alpha)]$ in
equation (16) is close to unity. We can see from Figure \ref{fig:alpha5pc}
that $p(\alpha)$
is somewhat flatter in this range than for $\alpha \simeq 1$, and this
behaviour is reflected in the shape of the higher and lower order statistics
distributions.

Our aim is to use the properties of the $\alpha_{(r)}$ distributions to
define a better distance estimator. The question of how best to combine some
or all of the ordered $D \sin i$ values measured from our sample
into one single distance estimate is non-trivial, however. One could, for
example, derive a {\em separate} distance estimate - analogous to
$\rmsub{\hat{D}}{mean}$ or $\rmsub{\hat{D}}{med}$ - from each ordered
$D \sin i$ in turn,
and take the mean of these individual estimates as our adopted cluster
distance; the shapes of the order statistic distributions shown in Figure
\ref{fig:ordstats}
suggest that it would be inappropriate to assign equal weight to each order,
however.

Rather than, for example, adopting some {\em ad hoc} weighting scheme to
resolve this problem, we construct an `ordered' distance estimator,
$\rmsub{\hat{D}}{ord}$, which combines the measured $D \sin i$ values by a
different - and rather more elegant -
method: one which accounts naturally for the shape of each order statistic
distribution in its definition.
\begin{figure*}
  \centering
  \vbox to 165mm{}
\caption{Histograms of naive, mean and median cluster distance
estimator distributions, derived from 2000 samples of 30 stars at a true
distance of $200 \,$pc. Results are shown for
for an angular diameter error dispersion of $\sigma_{\phi} = 0.05$ and
$0.1$}
\label{fig:est30stars}
\end{figure*}
\subsection{`Ordered' Estimator, $\rmsub{\hat{D}}{ord}$}

Suppose we measure the value of $D \sin i_{(r)}$, for the $r^{th}$ star in our
ordered sample. Given this measured value, the probability distribution,
$p_{r}(\alpha_{(r)})$, of the $r^{th}$ order statistic of $\alpha$ might now
equivalently be regarded as a probability distribution, $\lambda_{r}(D)$,
for the true cluster distance, $D$, viz:-
\begin{equation}
\lambda_{r}(D) \equiv p_{r}(\alpha_{(r)} = \frac{D \sin i_{(r)}}{D} )
\end{equation}
This equation follows from equation (9) above. $\lambda_{r}(D)$ is referred to
as the {\em likelihood function} for $D$, given the measured value of
$D \sin i_{(r)}$. Let $\Lambda(D)$ denote the product of the likelihood
functions for {\em all} orders, $r = 1, \ldots , n$. viz:-
\begin{equation}
\Lambda(D) = \prod_{r=1}^{n} \lambda_{r}(D)
\end{equation}
We define our ordered cluster distance estimate, $\rmsub{\hat{D}}{ord}$, as the
value of $D$ which maximises $\Lambda(D)$; i.e. $\rmsub{\hat{D}}{ord}$
satisfies:-
\begin{equation}
\frac{\partial \Lambda}{\partial D} | _{D = \rmsub{\hat{D}}{ord}}
\hspace{1mm} = \hspace{1mm} 0
\end{equation}
Since we do not have an analytic form for $\Lambda(D)$ we cannot calculate
$\rmsub{\hat{D}}{ord}$ directly by differentiation. It is, nevertheless,
straightforward to determine $\rmsub{\hat{D}}{ord}$ simply by computing
$\Lambda(D)$ over a range of trial distances and finding the distance which
yields the maximum value.

It is instructive to compare the properties of $\rmsub{\hat{D}}{ord}$ with
the other
estimators which we have discussed, and discover if it is indeed a `better'
estimator. We can do this again by using Monte Carlo simulations to
construct the distribution of $\rmsub{\hat{D}}{ord}$ at a given true distance.

As an illustration Figure \ref{fig:ordest} shows the results obtained from
computing
$\rmsub{\hat{D}}{ord}$ for 2000 cluster samples at a true distance of
$200 \,$pc.
As in Figures \ref{fig:est10stars} to \ref{fig:est30stars}, sample sizes of
$10, 20$ and $30$ stars and
angular diameter errors of $5\% \, $ and $10\% \, $ are considered. A range
of trial
distances from $100 \,$pc to $300 \,$pc, at intervals of one pc, was tested.
Given the spread in the distributions obtained, clearly a smaller step-size
would have been redundant.

Figure \ref{fig:ordest} demonstrates that $\rmsub{\hat{D}}{ord}$ is a slightly
more accurate distance
estimator than each of three estimators previously considered: in all cases
the bias of $\rmsub{\hat{D}}{ord}$ is found to be negligible, and the
dispersion
appreciably smaller than that of both $\rmsub{\hat{D}}{mean}$ and
$\rmsub{\hat{D}}{med}$.
\begin{figure*}
  \centering
  \vbox to100mm{}
\caption{Distribution of the order statistics, $\alpha_{(r)}$, of the
  random variable $\alpha$, with pdf as given by Figure 1, from a sample
size of n = 10. The r = 1, r = 4, r = 7 and r = 10 order statistics are
shown.}
\label{fig:ordstats}
\end{figure*}
\begin{table*}
  \centering
\caption{Mean value and dispersion of naive, mean, median and
ordered cluster distance estimators, computed from 2000 simulations of
cluster samples at a true distance of $200 \,$pc. Results are given, in pc,
for a sample size of $n = 10$, $20$ and $30$ stars, and for an angular
diameter error dispersion of $\sigma_{\phi} = 0.05$ and $0.1$}
\label{tab:estdistrib}
\begin{flushleft}
{\large $\sigma_{\phi} = 0.05$}
\end{flushleft}
\vspace{-8mm}

\renewcommand\arraystretch{1.3}

\begin{tabular}[t]{lcllccllccll}

& &\multicolumn{2}{c}{n = 10} & & &
   \multicolumn{2}{c}{n = 20} & & &
   \multicolumn{2}{c}{n = 30} \\

$\hat{D}$ & &
{\small mean} & $\sigma_{\hat{D}}$ & & &
{\small mean} & $\sigma_{\hat{D}}$ & & &
{\small mean} & $\sigma_{\hat{D}}$ \\
\hline

$\hat{D}_{naive}$ & & 221.8 & 16.5 & & & 231.5 & 14.8 & & & 236.1 & 13.4 \\
$\hat{D}_{mean}$  & & 200.2 & 14.6 & & & 199.9 & 10.0 & & & 199.8 &  8.5 \\
$\hat{D}_{med}$   & & 199.4 & 16.2 & & & 199.8 & 11.6 & & & 199.8 &  9.8 \\
$\hat{D}_{ord}$   & & 199.9 & 12.2 & & & 199.9 &  8.5 & & & 199.9 &  7.1 \\
\end{tabular}
\vspace{1cm}

\begin{flushleft}
{\large $\sigma_{\phi} = 0.1$}
\end{flushleft}
\vspace{-8mm}

\renewcommand\arraystretch{1.3}

\begin{tabular}[t]{lcllccllccll}

& &\multicolumn{2}{c}{n = 10} & & &
   \multicolumn{2}{c}{n = 20} & & &
   \multicolumn{2}{c}{n = 30} \\

$\hat{D}$ & &
{\small mean} & $\sigma_{\hat{D}}$ & & &
{\small mean} & $\sigma_{\hat{D}}$ & & &
{\small mean} & $\sigma_{\hat{D}}$ \\
\hline

$\hat{D}_{naive}$ & & 232.8 & 24.2 & & & 246.3 & 22.9 & & & 253.3 & 21.3 \\
$\hat{D}_{mean}$  & & 200.3 & 15.9 & & & 199.8 & 10.7 & & & 199.7 &  9.2 \\
$\hat{D}_{med}$   & & 200.3 & 17.6 & & & 200.3 & 12.5 & & & 200.5 & 10.7 \\
$\hat{D}_{ord}$   & & 200.1 & 14.6 & & & 200.1 &  9.8 & & & 200.2 &  8.4 \\
\end{tabular}
\end{table*}
These results are also summarised in Table \ref{tab:estdistrib}. We can see
from Table \ref{tab:estdistrib} that the dispersion of $\rmsub{\hat{D}}{ord}$
is around $10\% \, $ smaller than that of $\rmsub{\hat{D}}{mean}$ in all
cases. The
improvement gained by using $\rmsub{\hat{D}}{ord}$ appears to be greater for
smaller sample sizes: i.e. there is less difference between
$\rmsub{\hat{D}}{ord}$ and
$\rmsub{\hat{D}}{mean}$ for $n = 30$ than for $n = 10$. This would seem to be
consistent with the central limit theorem, which requires that
$\rmsub{\hat{D}}{mean}$
is asymptotically normally distributed (and hence a sufficient
statistic) as n increases. It is also worth noting that the dispersion of all
of the estimators decreases more noticeably when one increases the sample
size from $n = 10$ to $n = 20$, as compared with an increase from $n = 20$
to $n = 30$. Further studies bear out this trend for larger samples: beyond a
sample size of $n \simeq 30$ there is very little further gain in the
accuracy of the distance estimators - and in particular
$\rmsub{\hat{D}}{ord}$ - by
adding new stars to the cluster sample.

The estimator dispersions reported in Table \ref{tab:estdistrib} provide a
direct measure of the relative accuracy of each distance estimator at a true
distance of $200 \,$pc. In particular, we find that $\rmsub{\hat{D}}{ord}$ is
found to be
accurate to $\sim 7\% \, $ (at the $1 \sigma$ level) in the worst case
examined
where we have only $10$ stars and with $\sigma_{\phi} = 0.1$; this accuracy
improves to slightly more than $4\% \, $ with a sample of $30$ stars. For
$\sigma_{\phi} = 0.05$ the accuracy of $\rmsub{\hat{D}}{ord}$ improves from
$\sim 6\% \, $ to $\sim 3.5\% \, $ as the sample size increases from $n = 10$
to $n = 30$.

\subsection{Properties of Distance Estimators: Summary}
The results of our Monte Carlo simulations - as summarised in Table
\ref{tab:estdistrib} and
illustrated in Figures \ref{fig:est10stars}, \ref{fig:est20stars},
\ref{fig:est30stars}, and \ref{fig:ordest} - demonstrate that the ordered
distance estimator, $\rmsub{\hat{D}}{ord}$, is the best of the four distance
estimators which we have considered in this paper. In all cases
$\rmsub{\hat{D}}{ord}$ is unbiased, and has the smallest dispersion.
\begin{figure*}
  \centering
  \vbox to 165mm{}
\caption{Histograms of ordered cluster distance
estimator distributions, derived from 2000 cluster samples at a true
distance of $200 \,$pc. Results are shown for a sample size of $10$, $20$ and
$30$ stars, and for an angular diameter error dispersion of
$\sigma_{\phi} = 0.05$ and $0.1$}
\label{fig:ordest}
\end{figure*}
The naive estimator, $\rmsub{\hat{D}}{naive}$, would clearly be a poor choice
on the
other hand. This estimator is positively biased, resulting from the naive model
for the distribution of $D \sin i$ which is adopted in its definition.
Moreover,
the bias of $\rmsub{\hat{D}}{naive}$ increases as the number of sampled stars
increases. This property is particularly undesirable, since in the
statistics literature one meets many biased estimators which are,
nevertheless, {\em consistent} - meaning that their bias tends asymptotically
to zero with increasing sample size. $\rmsub{\hat{D}}{naive}$ instead displays
precisely the opposite behaviour.

It is straightforward to compute the distributions of these estimators
at a range of different true cluster distances, and the same qualitative
results are found in all cases, thus consolidating our choice of
$\rmsub{\hat{D}}{ord}$ as best cluster distance estimator.

Moreover, this Monte Carlo
approach clearly provides a simple means of assigning errors to cluster
distance estimates derived from real data - as we have already indicated in
discussing the results of Table \ref{tab:estdistrib} above. Suppose, for
example, that we
infer an ordered distance estimate of $\rmsub{\hat{D}}{ord} = \Delta$ from a
given
sample of real data . To calculate an error estimate we first generate a large
number of random samples, each equal in size to the observed
sample and assuming $\rmsub{D}{true} = \Delta$, and from a histogram of the
distance estimates for these artificial samples we derive the distribution of
$\rmsub{\hat{D}}{ord}$ at this true distance. We then adopt the dispersion of
this distribution as our error estimate for the cluster distance.
In the same way we can use the distribution of $\rmsub{\hat{D}}{ord}$ (or any
other estimator) derived from our simulations to determine confidence
intervals for the true cluster distance, given our estimated value.

This approach is straightforward to implement, but does suffer from one
specific technical loophole: this concerns the fact that we generate the
distribution of our estimator assuming $\rmsub{D}{true} = \Delta$, when it is
not $\rmsub{D}{true}$ but rather our {\em estimated} distance,
$\rmsub{\hat{D}}{ord}$, which
is equal to $\Delta$. A more rigorous method for assigning errors and
determining confidence intervals which overcomes precisely this
loophole does exist, and is described in detail in e.g. \scite{HendrySimmons90}
or \scite{MoodGray74}.
In the present context, however, we find that the confidence intervals
derived by the simple method outlined above are essentially identical to
those derived by the more rigorous approach, and so we do not describe the
latter method here.

The smaller dispersion of $\rmsub{\hat{D}}{ord}$, as compared with
$\rmsub{\hat{D}}{mean}$
and $\rmsub{\hat{D}}{med}$, illustrates that by ordering the sampled
$D \sin i$ values
one can frequently define a better cluster distance estimator;
$\rmsub{\hat{D}}{ord}$
being an example of one such estimator. There are a number of other methods
by which one can use the properties of ordered samples to construct a cluster
distance estimate. One appealing technique is to model the cumulative
distribution function of $D \sin i$ for a cluster at a given true distance,
and then construct a sample cumulative distribution function to be
compared with the model distribution. One would adopt as the
cluster distance estimate the distance which `best fits' the modelled
$D \sin i$ distribution to the sampled distribution. A suitable criterion
for identifying the best fit might be, for example, the distance which
minimises the Kolmogorov-Smirnov statistic for the two cumulative
distributions. This statistic is frequently used in problems of this type and
is particularly robust to the form of the underlying distribution
(c.f. \scite{KendallStuart63}). We have investigated
the use of this method in the present context, however, and find that it
gives no better (and frequently worse) results than using
$\rmsub{\hat{D}}{ord}$.
The robustness of this approach would be particularly useful, however, if one's
model for the distribution of $D \sin i$ were in some way uncertain or
ambiguous, and we will investigate the use of more robust estimation
techniques in future work.

\section{BAYESIAN CLUSTER DISTANCE ESTIMATES}
\label{sec:bayes}

In this section we briefly consider the estimation of the cluster distance as a
problem in Bayesian inference. We will see that such a treatment is
complementary to the analysis of distance estimators developed in section
\ref{sec:ests}: indeed a Bayesian approach follows quite naturally from our
chosen form for the ordered distance estimator.

The basic elements which comprise a Bayesian treatment of the problem can be
summarised as follows. We begin by postulating a {\em prior} distribution for
the parameter which we wish to estimate - i.e. the true cluster distance,
$\rmsub{D}{true}$. This distribution is supposed to express our state of
knowledge
or ignorance about $\rmsub{D}{true}$ before any data are obtained - in our
case the
data being our ordered set of $D \sin i$ values for the sampled stars
in the cluster. Note that this postulate immediately represents a departure
from our earlier view that $\rmsub{D}{true}$ was a fixed, although unknown,
parameter; we are now choosing to regard both $D \sin i$ and
$\rmsub{D}{true}$ itself as random variables.
Next we introduce a model which describes the probability of observing the
data given the parameter value, $\rmsub{D}{true}$ - in other words the expected
distribution of ordered $D \sin i$ values in a sample drawn from a cluster at
a given true distance.

The essential idea of the Bayesian approach is to combine these two
distributions - our prior distribution for $\rmsub{D}{true}$, and the $D \sin
i$
distribution given $\rmsub{D}{true}$ - to derive a {\em posterior} distribution
for $\rmsub{D}{true}$, which expresses our state of knowledge of
$\rmsub{D}{true}$ {\em after} we have measured the values of $D \sin i$. The
form of the posterior
distribution is given by Bayes' theorem, and can be stated in the present
context as:-
\begin{equation}
p(\rmsub{D}{true} | data) \hspace{1mm} = \hspace{1mm} \kappa \;
p(data |\rmsub{D}{true})p(\rmsub{D}{true})
\end{equation}
Here $p(\rmsub{D}{true})$ denotes our prior distribution for the true cluster
distance,
and $\kappa$ is a normalisation constant which ensures that the posterior
distribution integrates to $1$.

The Bayesian formulation is not so far removed from the ideas which
underpin our definition of $\rmsub{\hat{D}}{ord}$ in Section \ref{sec:ests}.
The
concept introduced in equation (17) of a likelihood function, $\lambda_{(r)}$,
for $\rmsub{D}{true}$ given the measured value of $D \sin i_{(r)}$, has already
alluded to the fact that we could regard $\rmsub{D}{true}$ as a random
variable. The Bayesian
viewpoint extends this interpretation of the likelihood function to derive not
simply a point estimate but rather a probability distribution for
$\rmsub{D}{true}$,
in the light of the observed data. In the same way as in section
\ref{sec:ests}, then, we can regard the probability distribution,
$p(data|\rmsub{D}{true})$, on the right hand side of equation (20) as a
function
of $\rmsub{D}{true}$: i.e. a likelihood function, $L(\rmsub{D}{true}|data)$,
for $\rmsub{D}{true}$
given the measured $D \sin i$ values.
The likelihood function clearly plays a crucial role in determining our
posterior distribution: it is the function through which the observed data
modifies our prior knowledge of $\rmsub{D}{true}$, and can therefore be
thought of
as representing the information about $\rmsub{D}{true}$ which comes directly
from the data.

In the present context $L(\rmsub{D}{true}|data)$ is, therefore, given by the
joint
distribution of ordered $D \sin i$ values expected in a cluster at true
distance, $\rmsub{D}{true}$, viz:-
\begin{equation}
L(\rmsub{D}{true}|data) = p_{1 \ldots n}( D \sin i_{(1)}, \ldots ,
D \sin i_{(n)} | \rmsub{D}{true} )
\end{equation}
It follows, moreover, from equation (9) that we can rewrite this as:-
\begin{equation}
L(\rmsub{D}{true}|data) = p_{1 \ldots n}( \alpha_{(1)}, \ldots , \alpha_{(n)})
\end{equation}
Where, of course, $\alpha_{(r)} = \frac{D \sin i_{(r)}}{\rmsub{D}{true}}$,
for all
$r = 1, \ldots, n$, as before. A general expression for this joint
distribution is given in \scite{David81} in terms of the density function and
cumulative distribution of $\alpha$. It is important to note that this joint
distribution will {\em not} in
general be equal to $\Lambda(D)$, as defined in equation (18) above, however.
In other words
the joint likelihood function for all of the $D \sin i$ order statistics is
not in general equal to the product of the individual likelihood functions.
This is because the sample $D \sin i$ values are not independent of each other:
the measured value of $D \sin i_{(n)}$, for example, must have a bearing on the
distribution of $D \sin i_{(n-1)}$, since we now have the constraint that
$D \sin i_{(n-1)} \leq D \sin i_{(n)}$, and so on for the smaller order
statistics.

We now illustrate the application of this Bayesian approach to some typical
cluster samples. Figure \ref{fig:bayesrec} shows the posterior
distribution for $\rmsub{D}{true}$ obtained from an ordered sample of
$D \sin i$
values, computed for sample sizes ranging between 10 and 30 stars.
All samples were drawn from the simulated $D \sin i$ distribution of a
cluster at a distance of $200 \,$pc, and with $\sigma_{\phi} = 0.05$.
In each case the prior distribution for $\rmsub{D}{true}$ was taken to
be uniform within the range $150$ to $250 \,$pc.

We can see from Figure \ref{fig:bayesrec} that the Bayesian approach yields
results which are broadly consistent with the estimator distributions
derived from Monte Carlo sampling in Section \ref{sec:ests}. With a sample
size of $10$ stars, for example, we find that the posterior distribution
has a dispersion of $\sim 10 \,$pc, which is consistent with the error
estimates
derived for $\rmsub{\hat{D}}{ord}$ in Table \ref{tab:estdistrib}. One can,
of course,
also derive Bayesian confidence intervals for $\rmsub{D}{true}$ directly from
the posterior distribution.

It is clear from Figure \ref{fig:bayesrec} that the posterior distribution
becomes `sharper' as the
sample size increases - indicating a more accurate Bayesian estimation of the
true distance from larger samples, as one would expect. Moreover, we find that
there is little further reduction in the dispersion of the posterior when
one's sample contains more than $\sim 30$ stars - which is also analogous to
the results of Section \ref{sec:ests}.

One need not carry out our Bayesian reconstruction using all $n$ order
statistics: one can use any single order statistic, or any subset of the
full range of orders, with results which accord with the distributions of the
order statistics used.
The posterior distribution recovered from using only the smallest $D \sin i$
value, for example, is considerably poorer than those shown in Figure
\ref{fig:bayesrec}: this merely reflects the relatively large dispersion of
the $\alpha_{(1)}$ distribution - as can be seen in Figure \ref{fig:ordstats}
above. In general we find that - for a given sample - the posterior of
smallest dispersion is obtained by using the complete ordered sample of
$D \sin i$ values.

The choice of prior distribution in the application of Bayesian
methods is clearly very important. The use of a non-uniform prior
is the source of considerable controversy in the statistics literature,
since it is often alleged that such a prior prejudices one's results by,
as it were, forcing the data to say something different about
$\rmsub{D}{true}$ than the information which they in fact contain. In
recognition of this point,
a uniform prior seems to us to be a more appropriate choice. In this case it
is solely the properties of the likelihood function which determine the form
of the posterior distribution - and not some predisposed view of what the true
cluster distance `should' be.

Nevertheless, a powerful feature of the Bayesian approach is the fact that
one can take the posterior distribution recovered from a given cluster sample
and adopt this as a {\em new} prior, in order to determine an `updated'
posterior distribution in the light of new data - i.e. as $D \sin i$ values
are measured for additional stars in the cluster. We will investigate this
extension to our analysis in future work.

\section{CONCLUSIONS}
\label{sec:conc}

In this paper we have presented a new method for determining the distance to
nearby open clusters. The method is applicable to clusters containing
fast-rotating late type stars, whose rotation periods have been measured from
detecting the rotational modulation of surface inhomogeneities. The period of
each star is then combined with its projected rotational velocity and an
estimate of its angular diameter, inferred from the Barnes-Evans relation, to
form an estimate of the projected cluster distance, $D \sin i$.

We have shown how one may then combine the set of $D \sin i$ values inferred
from each star in the cluster sample to form an `ordered' distance estimator
of the true cluster distance. We have investigated the properties of this
estimator using Monte Carlo simulations of cluster samples, after careful
modelling of the intrinsic scatter in the Barnes-Evans relation and the
observational selection effects to which the samples are subject.
\begin{figure*}
  \centering
  \vbox to100mm{}
\caption{Examples of the posterior distribution for the true cluster
distance, derived by applying the bayesian method described in the text to
simulated cluster samples containing between $10$ and $30$ stars. In all cases
the simulated samples were drawn from a cluster at a true distance of
$200 \,$pc}
\label{fig:bayesrec}
\end{figure*}
We have shown how one may apply this distance method to real data samples,
in order to derive error estimates and confidence intervals for the true
cluster distance. We have also demonstrated that this new method is
amenable to a Bayesian analysis, and have again illustrated this approach
using artificially generated cluster samples.

We have found that, for realistic models of the random variables, the
accuracy of our distance method is between $\sim 3$ and $7\% \, $
(at the $1 \sigma$ level) - depending on the size of the cluster sample - which
is comparable with the precision of distance estimates obtained by ZAMS
fitting techniques. Since it is subject to a different set of systematic
errors and model assumptions than ZAMS fitting - and in particular does not
rely upon a zero-point calibration by a single cluster, usually the Hyades -
these results indicate that our new method can play a useful and important
role alongside more traditional cluster distance indicators in better
determining the local distance scale.

In a forthcoming paper, currently in preparation, we will apply our new
distance method to real data samples taken from the $\alpha$ Per and Pleiades
clusters, and explicitly compare the results of our method with those obtained
by the ZAMS procedure. Other possible future applications of our method
include a better calibration of the Barnes-Evans relation - and determination
of the inclination distribution - for fast rotators in young open clusters,
by comparing our results with the very accurate cluster distances soon to be
provided by the Hipparcos satellite.

\vspace{5mm}
{\sc Acknowledgments} \\
 \\
MAH would like to thank John Simmons and Robert Smith for useful discussions
at various stages of this work. The authors also thank the anonymous referee
for helpful comments and suggestions.

We acknowledge the use of the {\sc{starlink}} Microvax 3400 computer,
funded by a SERC grant to the Astronomy Centre at the University of
Sussex. During the course of this work MAH was supported by a SERC
research fellowship, MAO was funded by a SERC research studentship and
ACC was supported by a SERC advanced fellowship
at the University of Sussex.

\end{document}